\documentclass{raa}
\usepackage{natbib}
\usepackage{graphicx,times}

\renewcommand\baselinestretch{1.11}

 \renewcommand{\thepage}{16--\arabic{page}}
\begin{document}

\title{Preface:
Planning the scientific applications of the Five-hundred-met{e}r
Aperture Spherical radio Telescope }

\volnopage{{\bf 2019} Vol.~{\bf 19} No.~{\bf 2}, ~16(2pp)~
   {\small  doi: 10.1088/1674--4527/19/2/16}}
   \setcounter{page}{1}

\author{Di Li
\inst{1,2,3}
\and John M. Dickey
\inst{4}
\and Shu Liu
\inst{1,2}
}

\institute{National Astronomical Observatories, Chinese
Academy of Sciences, Beijing 100101, China; {\it dili@nao.cas.cn}\\
\and Key Laboratory of FAST, National Astronomical Observatories,
 Chinese Academy of Sciences, Beijing 100101, China\\
\and
School of Astronomy and Space Science, University of Chinese Academy of
Sciences, Beijing 101408, China\\
\and School of Natural Sciences, Private Bag 37, University of
Tasmania, Hobart, TAS, 7001, Australia\\
\vs\no
   {\small Received 2018 October 8;    accepted 2018 November 28}}

\abstract { The Five-hundred-met{e}r {A}perture {S}pherical radio
{T}elescope (FAST) is by far the largest telescope of any kind
ever built. FAST produced its first light in September 2016 and it
is now under commissioning, with normal operation to commence{ in}
late 2019. During testing and early science operation, FAST has
started making astronomical discoveries, particularly pulsars of
various kind{s}, including {millisecond pulsars}, binaries,
gamma-ray pulsars, etc. The papers in this {mini-volume} propose
ambitious observational projects to advance our knowledge of
astronomy, astrophysics and fundamental physics in many ways.
Although it may take FAST many years to achieve all the goals
explained in these papers, taken together they define a powerful
strategic vision for the next decade. \\}

\titlerunning{{\it D. Li et al}.: ~Planning the scientific applications of FAST}
\authorrunning{{\it D. Li et al}.: ~Planning the scientific applications of FAST}

\maketitle

In the last few years, scientific observations have begun on several
powerful new radio telescopes. In many ways, the
Five-hundred-met{e}r Aperture Spherical radio Telescope
(FAST) is the most impressive of all these new instruments, a list
that includes MeerKAT (Camilo et al. 2018), ASKAP (Johnston et al.
2007), MWA (Tingay et al. 2013), LOFAR (van Haarlem et al. 2013),
APERTIF (Oosterloo et al. 2009) and the JVLA (Perley et al.
2011). All of these new instruments represent major investments in
scientific infrastructure that also drive improvements in industrial
and information technologies. As these telescopes
are completed and commissioned, opportunities for new kinds of
astronomical observations {similarly }open new doors
and windows leading to new landscapes.

Astronomers have not been waiting idly for FAST and the other new
telescopes to be completed.  Planning new
observations to find new {targets} and new kinds of
objects beyond the reach of existing facilities is one of the most
exciting jobs of a professional astronomer. Opportunities to attempt unprecedented
projects and observations, that go beyond anything that has been
done before, come only once in the careers of most astronomers.  The
papers in this {mini-volume} of Research in
Astronomy and Astrophysics are reports of planning exercises by
international teams of radio astronomers focused on how best
to apply the unprecedented power of the FAST radio telescope. Now
that commissioning is complete and the telescope has begun its first
major survey, these papers discussing the possibilities of the new
instrument have immediate importance.

\begin{table}

\centering
\begin{minipage}{5.2cm}
\caption{ FAST Technical Specifications \label{tab:FAST_specs}}
\end{minipage}

 \renewcommand\baselinestretch{1.3}
\fns\tabcolsep 2mm
\begin{tabular}{lp{2.8cm}}
\hline\noalign{\smallskip}
Parameter & Value \\
\hline\noalign{\smallskip}
Aperture diameter & 500{\,}m 
\\
Radius of curvature & 300{\,}m\\
Illumination diameter & 300{\,}m \\
Focal ratio $f/D$ & 0.461 \\
Latitude & +25\dg\ 39\arcmin \  11\arcsec \\
Longitude & 106\dg\ 51\arcmin \  24\arcsec \  E \\
Zenith angle range & 26.4\dg\ (with full gain), 40\dg\ (with reduced gain)\\
Frequency coverage & 70{\,}MHz--3{\,}GHz \\
Maximum slew time & 10{\,}min\\
Multi-beam receiver beams & 19\\
Multi-beam receiver frequency range & 1.04--1.45{\,}GHz\\
Multi-beam receiver sensitivity  ($T_{\rm rec}$) & 20{\,}K\\
Multi-beam receiver beam width & 2.9\arcmin \\
\hline
\end{tabular}
\end{table}

The FAST telescope's specifications and current status are
summari{z}ed {in} Table~\ref{tab:FAST_specs}, with more details to
be found in Li et al. (2018). The first major sky survey, the
Commensal Radio Astronomy FAST Survey, CRAFTS, is being planned
and its mode tested, using the FAST L-band Array of Nineteen-beams
(FLAN) receiver. CRAFTS aims to utilize 100\% of FAST's gain in
drift-scan mode, which minimizes the complexity and overhead in
pointing and tracking with more than 4000 primary panels and the
focal-cabin driven by {six} cables.  The key innovation of CRAFTS
is its capability of simultaneously recording pulsar, HI galax{y},
HI image, and {fast radio burst (}FRB{)} (transients) data
streams. The related technologies and calibration modes are being
demonstrated now and will be published soon. Deeper surveys are
also being planned, in particular, those of the Galactic plane and
M31 systems. These plans, along with more receivers and other
improvements, are pending.

Discoveries such as new pulsars and unknown structure in the
interstellar gas of the Milky Way have already been made, with more
coming day by day.  The capabilities of the instrument are living up
to the expectations of the authors of the papers in this
{mini-volume}. Most of the objectives
described here will be achieved in the coming decade, many in the
next two or three years.

The papers in this {mini-}volume cover a broad range of science
topics, starting with
\begin{itemize}
\item[--] the fundamental constants of nature (Chen et al.
2019){;}
\item[--] cosmic rays (James et al. 2019){;}
\item[--] exoplanets (Zarka et al. 2019){;}
\item[--] gravitational radiation (Hobbs et al. 2019){;}
\item[--] pulsar magnetospheres (Wang et al. 2019){;}
\item[--] thermodynamic phases of the interstellar medium (Heiles et al.
2019){;}
\item[--] interstellar masers (Zhang et al. 2019){.}
\end{itemize}

The authors are very ambitious. Given the excellent progress of FAST
commissioning and its expected transition into normal operation in
2019, such optimism is justified. Related to and also beyond the
scope of this {mini-}volume, there
{has} been growing interest in the
science potential of FAST.  Yu et al. (2017) predicted more than
{10\,000} HI absorption systems {can} be
discovered by FAST, thus adding a rare, valuable probe of
cold baryons. In addition to this mini-volume, an ensemble of recent publications also shed insights into the potential of FAST in various frontiers of radio astronomy. Zhang et al. (2018) looked into FAST's potential
of detecting{ the} radio afterglow of {gamma-ray
bursts}. Zhang (2018) emphasized FAST's
unique capability of catching remote FRBs to probe{ the}
inter-galactic medium. Liu et al. (2018) simulated pulsar detection
by FAST and estimated more than 1500 new detections with more than
200 new {millisecond pulsars}. These are just some
examples, many of which are also published {in} RAA.
{I}f all these proposed observations are
successful{,} the effects will advance, even revolutionize, many
areas of physics and astrophysics.

More discoveries will be made by the FAST telescope, beyond the
expectations described in these papers.  When such a powerful new
telescope begins its scientific observations, unexpected signals and
effects often emerge.  {However,} unexpected
discoveries can only be made when scientists are trying to do
difficult projects that push the new instrument to its limits of
sensitivity and versatility. The authors of the papers in this
volume have plans for difficult and demanding observations, going
beyond what has been done by other telescopes in the past. Doing
these observations may lead to serendipitous discoveries, because of
their ambitious and challenging performance requirements.  As these projects are begun over the next few years, FAST
will have an impact on {many areas of} astronomy and
astrophysics {around the world}. We cannot know everything
it will discover, but these papers describe how FAST may profoundly
change our understanding of the universe.

\begin{acknowledgements}

We are very grateful to the anonymous referees, whose comments and
suggestions led to major improvements in these papers.  The
authors had to put up with delays in the publication of this
{mini-volume}, and we are grateful for their
patience. The editors would like to acknowledge the support from the National Key R\&D Program of China (2017YFA0402600) and the \nsfc  (11725313).
\end{acknowledgements}


\begin{thebibliography}{99}
\small \setlength{\itemindent}{-3mm} \setlength{\itemsep}{-0.7mm}
\setlength{\baselineskip}{4.4mm}


\bibitem[Camilo et al.(2018)]{2018ApJ...856..180C}
Camilo, F., Scholz, P., Serylak, M., et al.\ 2018, \apj, 856, 180
\bibitem[]{c}
Chen, X.,   Ellingsen, S. P., \&   Mei, Y.
 2019, \raa, 19, 18 (arXiv:1904.03871)


\bibitem[]{h}
Heiles, C.,
 Li, D.,
     McClure-Griffiths, N., et al.
    2019, \raa, 19, 17 (arXiv:1904.01237)


\bibitem[]{h}
Hobbs, G.,
 Dai, S.,
   Manchester, R. N., et al. 2019,
\raa, 19, 20 (arXiv:1407.0435H)


\bibitem[]{j}
James, C. W., Bray, J. D., \& Ekers, R. D.,  2019, \raa, 19, 19 (arXiv:1608.02407)
\bibitem[Johnston et al.(2007)]{2007PASA...24..174J} Johnston, S., Bailes, M., Bartel, N., et al.\ 2007, \pasa, 24, 174
\bibitem[Li et al.(2018)]{2018IMMag..19..112L} Li, D., Wang, P., Qian, L., et al.\ 2018, IEEE Microwave Magazine, 19, 112
\bibitem[Liu et al.(2018)]{2018ProgAstro...36..2L} Liu, P., Wang, P., Li, D., et al.\ 2018, Progress in Astronomy, 36, 2

\bibitem[Oosterloo et al.(2009)]{2009wska.confE..70O}
Oosterloo, T., Verheijen, M.~A.~W., van Cappellen, W., et al.\
2009, Wide Field Astronomy \& Technology for the Square Kilometre
Array (SKADS 2009), 70 {\it http://pos.sissa.it/132/070/pdf}

\bibitem[Perley et al.(2011)]{2011ApJ...739L...1P}
Perley, R.~A., Chandler, C.~J., Butler, B.~J., \& Wrobel, J.~M.\ 2011, \apjl, 739, L1
\bibitem[Tingay et al.(2013)]{2013PASA...30....7T} Tingay, S.~J., Goeke, R., Bowman, J.~D., et al.\ 2013, \pasa, 30, e007

\bibitem[]{}
Wang, H. G., Qiao, G. J., Du, Y. J., et al.,  2019, \raa, 19, 21

\bibitem[van Haarlem et al.(2013)]{2013A&A...556A...2V} van Haarlem, M.~P., Wise, M.~W., Gunst, A.~W., et al.\ 2013, \aap, 556, A2
\bibitem[Yu et al.(2017)]{2017RAA....17...49Y} Yu, H.-R., Pen, U.-L., Zhang, T.-J., Li, D., \& Chen, X.\ 2017, \raa, 17, 49

\bibitem[]{}
Zarka, P., Li, D., Grie{\ss} meier, J.-M.,   et al.,  2019, \raa,
19, 23 (arXiv:1904.01239 )


\bibitem[Zhang(2018)]{2018ApJ...867L..21Z} Zhang, B.\ 2018, \apjl, 867, L21

\bibitem[]{}
Zhang, J. S., Li, D., Wang, J. Z., et al., 2019, \raa, 19, 22

\bibitem[Zhang et al.(2018)]{2018ApJ...865...82Z} Zhang, Z.~B.,
Chandra, P., Huang, Y.~F., \& Li, D.\ 2018, \apj, 865, 82

\end{thebibliography}
\end{document}